# Inclusive Learning Analytics with Embedded Data Comics: A Conceptual Framework for Public Understanding of AI Ethics

Mengyi Wei [1]*, Chenyu Zuo[2], Dongsheng Chen[1], Liqiu Meng[1]

[1]*Chair of Cartography and Visual Analytics, Technical University of Munich*

[2] *Department of Civil, Environmental and Geomatic Engineering, ETH Zurich*

**ABSTRACT**

Public awareness of AI ethics plays a crucial role in fostering the responsible and sustainable development of AI technology. However, finding effective ways to promote public understanding of the ethical risks of AI remains a challenge. Given the complexity of AI ethical issues and the cognitive limitations of the public, this review paper proposes a conceptual framework for inclusive learning analytics with embedded data comics. Data comics help transform complex and abstract AI ethics cases into compelling and relatable stories, fostering public empathy and introspection. More importantly, inclusive learning analytics targets not only people of different demographic attributes, but also different mindsets with inherent cognitive biases. By providing equal and easily accessible channels for AI ethics issues, we aim to encourage the public to reflect on AI ethics incidents from multiple perspectives and develop the habit of continuous learning to adapt to evolving AI technologies and ethical risks.
**Keywords:** AI Ethics, News Reports, Narrative Visualization, Data Comics, Inclusive Learning Analytics

## 1 INTRODUCTION

AI models, led by LLMs such as ChatGPT and Deepseek, have brought tremendous convenience to society and are increasingly penetrating all aspects of social life. While there is a growing global interest in further enhancing the various AI capabilities, including linguistic, computational, and even spatial capabilities, the issue of AI ethics has not yet received sufficient attention. For example, our society increasingly relies on predictive models for credit scoring, automated screening of job applicants, and police analysis of potential suspects (Žliobaitė 2017), and these data-driven decisions are often implicitly racialized and gendered (Arnold et al. 2021). Furthermore, AI is prone to be used by capital to gain insights into the psychological state of the user, thereby manipulating public behavior (Belk 2021). Most people are unaware that they are living in the midst of AI risks. As our society focuses on how to make AI as intelligent as humans, we may be overlooking a crucial question: how should humanity respond to this unprecedented technological revolution (Box-Steffensmeier et al. 2022)? Harari (2016) wrote in his book, "*what will happen to society, politics, and daily life when unconscious but brilliant algorithms know us better than we know ourselves*?"

AI ethical issues refer to the moral challenges in the development, deployment, and use of AI. There is an urgent need to raise awareness of AI ethics (Dabis and Csáki 2024) because low public awareness of AI ethics can easily trigger deep-rooted social risks, such as the invisible erosion of individual rights and the absence of a foundation for regulatory public opinion (Douglas et al. 2024). However, the complexity of AI ethics and the limitations of the public's knowledge make it difficult to raise awareness. The complexity of AI ethics extends beyond technical considerations, encompassing interconnected social, legal, economic, and philosophical dimensions (Wei et al.

2025). Real-world AI ethics issues are often highly complex, ambiguous, and dynamic. Different stakeholders are involved, including governments, businesses, developers, and users, with varying levels of knowledge, expectations, personal interests, values, and goals. For example, businesses may prioritize the commercial value of AI, while governments focus on regulation and security, and ordinary users are more concerned with privacy and fairness. These conflicting interests make it challenging to establish a unified solution. In addition, conflicts and contradictions in solutions tend to be dynamic rather than universally fixed or permanent (Head and Alford 2015; Head 2022). Recognizing this characteristic is crucial to understanding that policies addressing AI ethics can never be optimal in an engineering sense; instead, they should be designed as effective policies incorporating insights from stakeholder participation. On the other hand, the limitations of public's knowledge often shape the understanding of AI ethics, with individuals from different backgrounds exhibiting varying cognitive biases. For example, some fear the changes brought about by AI out of ignorance, while others are overconfident in their own views. Overcoming these cognitive biases is not trivial, but necessary, for effective participation in the development of AI technology.

This paper introduces a framework of inclusive learning analytics with embedded data comics (**ILA-DC**), which focuses on promoting public awareness of AI ethics in an engaging and fair manner (see related definitions in Box 1). We explored how data comics embedded in inclusive learning analytics can address the complexity of AI ethics and break through public cognitive limitations. Through real-world AI ethics cases, we intend to highlight the potential of this framework in enhancing public understanding of AI ethics. As AI technologies advance rapidly, this is a pivotal moment to enhance public adaptability, strengthen ethical awareness, reflect on behavioral habits, and sustain lifelong learning, so we can actively shape the future rather than passively receive it.

Box 1: Definitions of the concepts for enhancing public awareness of AI ethics.

| |
|---|
| • **AI ethical issues:** The moral challenges in the development, deployment, and use of AI. These challenges often relate to the behavior of AI systems, the transparency of decision-making, fairness, security, privacy, and attribution of responsibility. The highly cross-cutting nature of AI ethics issues requires the participation of multiple stakeholders, including developers, policymakers, users, academics, and the public. All stakeholders must develop and implement policies that align AI with society's ethical values to ensure that AI truly serves humanity. |
| • **Inclusiveness:** It does not merely refer to care for disadvantaged groups. Rather, it denotes a form of cognitive openness that emphasizes diversity and the coexistence of differences. It concerns how systems can accommodate and respond to diverse cognitive styles, cultural backgrounds, and value orientations at both the levels of learning content and learners. |
| • **Learning analytics:** It is no longer a quantitative analysis of individual learning behaviors and educational performance in the traditional sense, but is instead embedded within a more social and reflective framework. It focuses on cognitive learning at the public level and the ethical mechanisms within sociocultural contexts, emphasizing data-driven learning reflection and public enlightenment. |
| • **Inclusive learning analytics:** It refers to both content inclusivity (translating complex AI ethics knowledge into data comics that are easier to understand) and audience inclusivity (accommodating learners from diverse cognitive backgrounds), while further optimizing the learning process by analyzing users' pathways of understanding. |
| • **Media communication:** The process of disseminating information, ideas, or stories through the media such as radio, television, newspapers, Internet.
• **News Reports:** The communication process of actual events containing facts in text, images, audio, or video.
• **Narrative visualization:** A method of storytelling with data through visual elements (such as charts, illustrations, animations, etc.) that are engaging and memorial. It aims to guide the viewer through complex content in a sequence of visualization steps.
• **Data comics:** A method of radical storytelling with data using visually appealing comic strips that contain elements such |

> as characters, plots, and dialogs.
>
> - **Public awareness:** It refers to the understanding, views, attitudes, and beliefs of social groups towards specific things, phenomena, issues, or concepts. It reflects not only the knowledge and viewpoints of individuals but also collective thought patterns and societal consensus. It is often influenced by various factors, such as information dissemination, personal experience, cultural background, and social environment.
> - **Public awareness of AI ethics:** It refers to the understanding, views, attitudes, and beliefs of social groups regarding the ethical issues, moral standards, and social impacts associated with artificial intelligence technology and its applications. It includes recognition of ethical topics such as the potential risks, responsibilities, privacy protection, fairness, and transparency of AI technology, as well as the public's assessment of how these issues apply and impact real-life situations.

## 2 THE ILA-DC FRAMEWORK

### 2.1 Inclusive Learning Analytics

Enhancing public awareness of AI ethics across diverse backgrounds is no easy task. The rapidly evolving big data and analytical technologies have led to the development of learning analytics (Ferguson 2012). Learning analytics is a product of the AI era, leveraging technological tools to develop learning platforms and tools that meet evolving needs. Meanwhile, it records and analyzes learning data to enable iterative updates, continuously optimizing the learning process. While current applications of learning analytics are primarily concentrated in higher education to enhance academic learning or professional skill development, little research has focused on the general public, especially those from varied backgrounds. The diversity in public knowledge and experience presents a significant challenge in identifying suitable approaches to address their learning needs.

To address the research gap, we propose a novel approach of inclusive learning analytics, which builds upon learning analytics while emphasizing two additional aspects: 1) Inclusivity of learning content – to ensure that the learning content is easily accessible and appealing. 2) Inclusivity of learners – to meet the changing needs of groups with diverse backgrounds (Figure 1). Content inclusivity involves transforming complex knowledge into comprehensible and engaging information, encouraging broader participation and interest. Learner inclusivity goes beyond the relatively static demographic attributes by emphasizing habits and knowledge frameworks that include different degrees and types of bias. Our approach focuses on behavioral diversity, aiming to ensure that all behavioral patterns of learners (the public) are understood and accepted rather than favoring a specific behavior type. The goal of inclusive learning analytics is to provide individuals with fair learning opportunities to understand AI risks, ultimately enabling them to identify suitable strategies to overcome their cognitive blind spots when developing or using AI products or services.

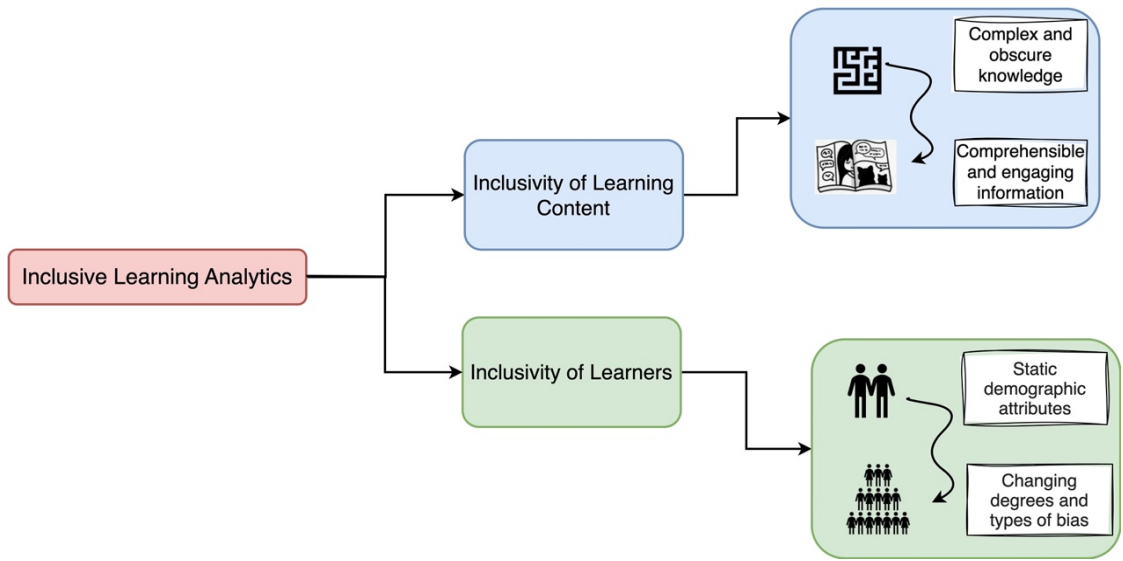

Figure 1. Two aspects of inclusive learning analytics

**2.2 Data Comics**

Applying inclusive learning analytics to enhance public awareness of AI ethics requires an appropriate medium. As narrative beings, humans tend to make sense of their lives, construct their worlds, and connect with others through storytelling (Frank and Seale 1996; Baldwin 2005). Data comics are a unique form of narrative visualization, which combines data visualization and comic narratives to present and explain data through comic devices such as characters, plots, and dialogues, presenting complex data and information to audiences in a storytelling way (Zhao et al. 2015; Bach et al. 2017; Wang 2022).

Research has shown that the effective use of text and visuals can enhance the comprehension of complex information, especially for individuals with limited education and for marginalized groups (Green and Myers 2010; Ahmed-Husain and Dunsmuir 2014; Al-Jawad and Frost 2014; Engebretsen et al. 2020). Data comics with their easily understandable and appealing format serve non-specialist viewers very well. They use panels as the core unit to encapsulate specific information in an integrated combination of text and images (Scott 2006). The panels help break down the data into multiple levels of detail, presenting them progressively through a split-screen design, making this information more likely to grab the reader's attention and be easily assimilated (Wang et al. 2019). In addition, data comics incorporate humorous elements, anthropomorphic characters, or narrative plots to make serious data issues lively and interesting. This can not only attract viewers' attention, but also ease their resistance to data and complex topics, thus increasing acceptance.

Unlike other forms of narrative visualization, data comics do not aim to offer complete solutions. Instead, they are designed to evoke emotional engagement and encourage deeper reflection among readers (Engebretsen et al. 2020). Narrative persuasion has been demonstrated in research by a wide range of scholars (Busselle and Bilandzic 2008; Graaf et al. 2016). Research has shown that personal stories or narratives, whether communicated face-to-face, in writing, or online, can motivate individuals to take action and support government action on important social issues (Polletta and Redman 2020). Data comics, use characters, plots, and dialogue to inject elements of emotion and context to data and integrate them into specific storylines. In this way, readers are brought closer to the story's protagonist and can easily relate to their own lives, thus triggering emotional resonance and self-reflection. In addition, data comics interpret the same theme from multiple perspectives through different characters, scenarios, or data levels. This multi-perspective presentation enables readers to examine issues from different angles, overcoming their cognitive limitations, and triggering deep thoughts. These features make data comics an ideal presentation form for telling AI ethics stories. Figure 2 shows the same event in two forms: an objective map (Figure 2a) and a subjective comic (Figure 2b). Both depict Cape Hellas and the extent of the British advance at the end of the campaign. The map in Figure 2a is factual, showing a rational and objective marching route. In contrast, Figure 2b combines comic characters with the map, adding humor, analogy, and sarcasm to convey the atmosphere of the scenario. This comic approach is more likely to evoke emotional resonance with the audience.

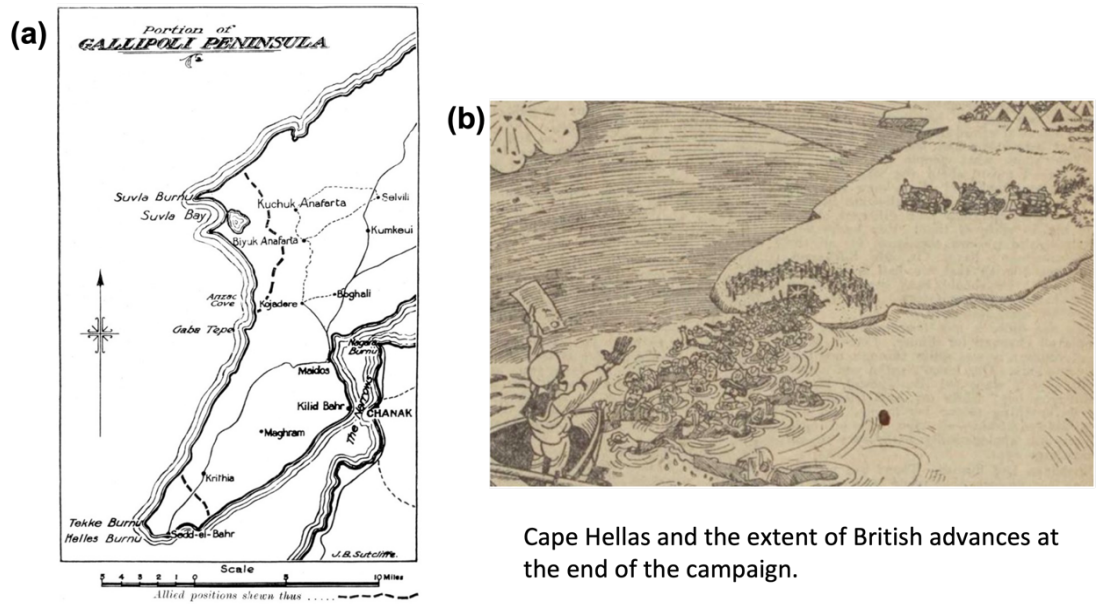

Figure 2. Cape Hellas and the extent of British advances at the end of the campaign (Moore and Cartwright 2014).

**2.3 The conceptual framework**

We propose to combine data comics with inclusive learning analytics in a new conceptual framework (ILA-DC) (Figure 3) for two main reasons. First, the specific learning barrier associated with complex AI ethics knowledge makes it difficult for the general public to understand the risks posed by AI technology. By combining data comics with a dashboard interface, real-world AI ethics events can be presented as engaging narratives. This can strengthen the connection between users' lived experience and digital data, make the latter more approachable, accessible, and visually appealing, thereby lowering the learning barrier for complex AI ethics knowledge. In this context, data comics provide not only effective learning opportunities for users with biased knowledge of AI, but also allow for greater inclusion of emotion and personalization (Williams 2013; Czerwiec et al. 2020), thus attract more people to participate due to the fun elements and power of humor to demystify complex statistics (Cole 2006; Alamalhodaei et al. 2020). Second, biases, be caused by ignorance or arrogance, can hinder the public's ability to effectively exercise its role as a voice in the development of AI technology. Data comics use character development and narrative design to present different perspectives that reflect the complexity and diversity of ethical issues. Personalized data comic panels can be designed to provide multiple learning paths, such as skipping mastered content or delving deeper into areas of uncertainty, based on learners' choices and feedback, so as to meet the needs of different groups, help break down their cognitive limitations, and improve the overall AI ethics literacy.

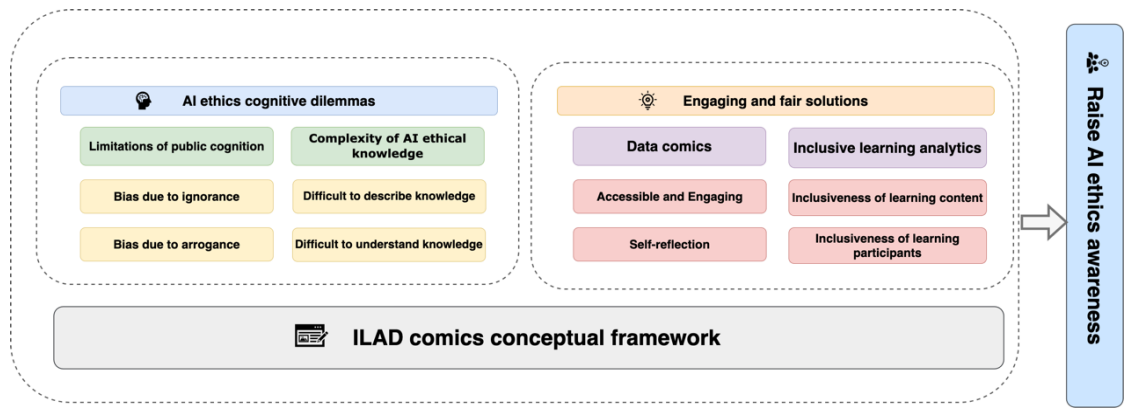

Figure 3. The conceptual framework of inclusive learning analytics for studying AI incidents.

## 3 POTENTIALS OF THE ILA-DC

**3.1 Data comics for promoting dynamic reflection on the real-world AI ethics.**

News of real-world AI ethics cases serve to challenge techno-utopian narratives and facilitate the materialization of legal and ethical governance structures. Currently, mainstream AI ethics narratives are primarily shaped by science fiction films and novels, as well as corporate marketing from major technology companies. These media sources significantly influence public perceptions of AI ethics, often raising questions about whether machines have developed human-like consciousness. Fictional portrayals of AI can distort public understanding by becoming a dominant lens through which people assess the technology, leading to unnecessary fears and misconceptions (Hermann 2020). Moreover, the emphasis on imagined AI scenarios in science fiction may cause the public to overlook urgent and everyday ethical issues related to AI in the real world, thereby missing critical opportunities for meaningful discussion and better ethical policymaking. Our ILA-DC framework uses real-world AI ethical events reported in the news media as a data source for the creation of comics. News media, known for its timeliness, accuracy, and objectivity, remains one of the primary ways people access information and understand facts. By focusing on actual AI ethics cases, data comics created under this framework aim to address real-world ethical challenges. This approach can support the development of an informed public understanding of AI ethics and foster more constructive discussions around policy and regulation.

In addition, presenting AI ethics events in the form of data comics fosters profound reader engagement and critical reflection through unresolved tensions and an ongoing space for discussion. Ethical issues in artificial intelligence are often highly complex, uncertain, and contested, involving different stakeholders with varying interests. Therefore, the conflicts and contradictions within potential solutions are usually dynamic rather than fixed or permanent (Machado and Silva 2025). Data comics, as a highly visual form of storytelling, can employ techniques such as ambiguous imagery, metaphorical expression, and open-ended dialog to introduce a sense of openness in their narratives. These features allow space for readers to explore and reflect on their own, rather than offering a definitive answer. As such, the ILA-DC framework holds advantages in addressing complex issues. It can support raising public awareness of AI ethics by encouraging discussion and debate to find a dynamic point of equilibrium.

**3.2 Inclusive learning analytics for cognitive debiasing.**

Bias due to ignorance - The development of AI technology cannot ignore non-experts' involvement (the less knowledgeable public), just as everyone's transportation choices are influenced by their cars, whether they own them or not. During AI development, the public with lower cognitive levels tends to form a collective imagination due to their limited understanding of the technology. This lack of understanding exacerbates anxiety about AI technology. Additionally, it deprives them of a voice in advocating for equal and fair AI development (Sartori and Bocca 2022). For example, in 2024, a Chinese tech giant company launched a self-driving car, the APOLLO Go, in Wuhan, to meet public travel needs. Self-driving cars alleviated the public's difficulty in getting a taxi during peak hours and were popular because they were standardized, safe, and inexpensive. However, the phenomenon has also greatly impacted the traditional service industry, causing strong resistance from drivers[1]. Apart from emotional relief and policy protection (Cath 2018; Russell 2021), the government should play a role in promoting new technology. The reaction of human drivers is the true epitome of non-experts' attitude toward AI technology. Due to cognitive limitations, it is difficult for non-experts to familiarize themselves with AI technology and to recognize the real risks that the technology poses. Instead, they strongly resist the technology because of the imagined risks, failing to see the new opportunities and impeding the sustainable development of AI technology.

Bias due to arrogance - Social psychology suggests that people tend to be influenced by pre-existing beliefs when seeking information (Johnston 1996), preferring supportive information over conflicting information (Jonas et al. 2001). More seriously, people may become trapped in an information cocoon when they seek only supportive or comforting information (Sunstein 2006; Peng and Liu 2021). This more knowledgeable segment of the population is also highly susceptible to information bias when confronted with AI ethical issues. For example, the work of AI development engineers mainly involves designing, developing, and optimizing AI models that can perform complex tasks, which leads them to place higher priority on the efficiency and application of models at the expense of ethical rules such as fairness (Umbrello 2022). Especially with rapid development of AI technology, this kind of entrenched mindset not only easily leads to professional stagnation, but also easily ignores the needs of different groups and fails to realize the sustainable development of the technology. Therefore, while technology iterates and innovates, people with high cognitive levels must also open their minds to deal with conflicting information, maintain the habit of lifelong learning and growth.

**3.3 ILA-DC framework as a bridge between individual diversity and collective understanding**

In the context of enhancing public awareness of AI ethics, inclusive learning analytics with embedded data comics exemplifies two complementary cognitive mechanisms: the preservation of individual diversity and the construction of collective understanding (Huberman and Glance 1993; Gardner and Garr-Schultz 2017; Tripathi 2019). Data comics embody the value of individual diversity by engaging audiences through narrative forms that prioritize personal empathy, emotional identity, and interpretive openness. Their storytelling flexibility accommodates readers from diverse social and cultural backgrounds, allowing everyone to engage with ethical dilemmas through their own life experiences and sense-making processes. By avoiding authoritative conclusions and offering multiple entry points to understand, data comics hold a diverse model of ethical education that empowers rather than instructs. In contrast, inclusive learning analytics focuses on the formation of collective understanding. By analyzing the learning behaviors and patterns of different populations, this approach identifies cognitive commonalities, gaps in understanding, and value mismatches. It enables the design of inclusive learning analytics systems that support shared ethical literacy while accommodating diversity in access and capability. In this sense, inclusive learning analytics help aggregate individual trajectories into a dynamic collective learning process. When integrated, these two approaches not only respect the subjectivity and variability of ethical interpretation but also facilitate the alignment of societal norms and shared values around AI ethics. The ILA-DC framework, combining two approaches, aims to achieve consensus within diversity. That is, while everyone retains the freedom to form their understanding, a certain degree of collective consensus is formed on key societal issues, thereby facilitating the development of AI ethics policies, and promoting the advancement of AI technology.

---

[1] https://www.sohu.com/a/792936280_121117078

**3.4 Empirical grounding from related domains**

Although the proposed framework of inclusive learning analytics with embedded data comics (ILA-DC) is conceptual, its feasibility is supported by empirical evidence from multiple related domains. Empirical studies also provide direct support for the effectiveness of data comics in delivering analytic feedback. In health communication, for instance, Fogwill and Manataki (2024) developed interactive data comics for public audiences to communicate cancer statistics and conducted a between-subject experiment with 98 participants comparing data comics with a text-based medium containing equivalent information. The results showed higher perceived engagement for data comics, and participants also demonstrated better recall and comprehension of the data in the narrative condition than in the text condition. Such comparative evidence suggests that data comics are not merely a more visually appealing "wrapper," but can yield measurable gains in learning outcomes (e.g., understanding and memory), thereby providing an empirical basis for presenting learning analytics results in the ILA-DC framework through narrative and stepwise structures.

This paper takes a different approach to inclusive learning analytics. It moves away from older ideas of "inclusivity" and focuses on population diversity, including people with different knowledge and thinking styles. Previous research in learning analytics has also aimed for equity, diversity, and inclusion and stresses the importance of including the perspectives and needs of all stakeholders (students, instructors, and administrators) in both system design and evaluation. Involving these groups helps systems find out which information is easy to understand for different users and which explanations may cause confusion or distrust. The system can then adjust explanations, feedback, and interface features to ensure that analytic results are clear to people from diverse backgrounds. Research shows that inclusive learning analytics should not just show the same results to everyone. Instead, they should analyze how people with different cognitive backgrounds interpret information. By tracking users' focus points, reasoning, and explanation preferences, designers can improve feedback and learning support, making learning analytics more useful for everyone (Bayer et al. 2024).

**4. WORKING PRINCIPLE OF THE ILA-DC FRAMEWORK**

The conceptual framework of ILA-DC is developed with the aim to build a data comics library by collecting real-world AI ethics incidents. The ILA-DC framework does not simply convert a single AI ethics incident into a data comic; instead, it integrates inclusive learning analytics to construct a meaningful and accessible data comics library. The implementation strategies are illustrated in Figure 4, we use cognitive deconstruction to correct public cognitive biases by exploring multiple perspectives on a single event, multiple events from a single perspective, and multiple perspectives on multiple events. For example, in autonomous driving field, a single event from multiple perspectives can be illustrated by presenting the same incident from the viewpoints of pedestrian, car company, and driver. A single perspective across multiple events can be explored by selecting the car company's viewpoint to present different incidents. In multiple perspectives across multiple events, the first incident could be analyzed from the perspectives of police, driver, and lawyer. In contrast, the second incident could be examined from the viewpoints of passengers, car company, taxi driver, government, and technology researchers. The complexity of AI ethics issues often arises from conflicting interests among stakeholders, ambiguous boundaries, dynamic changes, and the absence of definitive solutions. Therefore, implementing an inclusive learning analytics strategy requires a systemic approach, collaborative governance, and adaptive adjustments to enable continuous exploration and iterative optimization. Throughout this process, inclusive learning analytics with embedded data comics serve to continually correct public cognitive biases, fostering a habit of lifelong learning and ultimately promoting harmonious coexistence between AI and humanity.

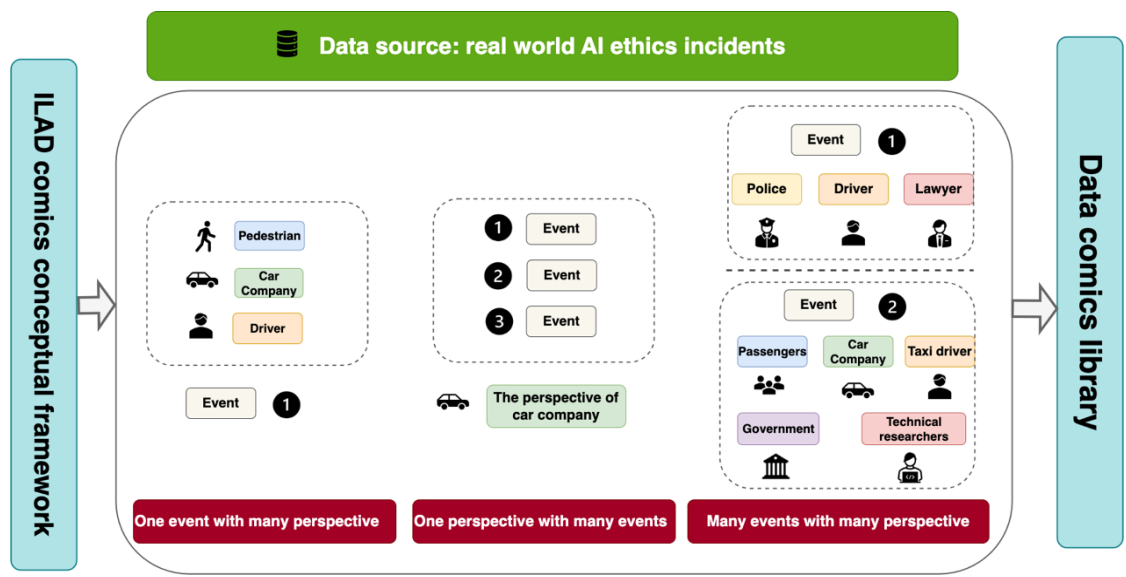

Figure 4. The implementation strategy of the inclusive learning analytics with embedded data comics for AI ethics incidents.

a. One event viewed from many perspectives for its inclusive understanding.
   **Incident Description:** An Uber vehicle was driving at night in autonomous mode when a pedestrian attempted to cross the road. Due to the limited skill level, the driver failed to brake in time, resulting in a collision that led to the pedestrian's death on the spot. The key stakeholders involved in this incident include the Uber driver, the pedestrian, and the autonomous vehicle company. They represent different perspectives as demonstrated in Figure 5.

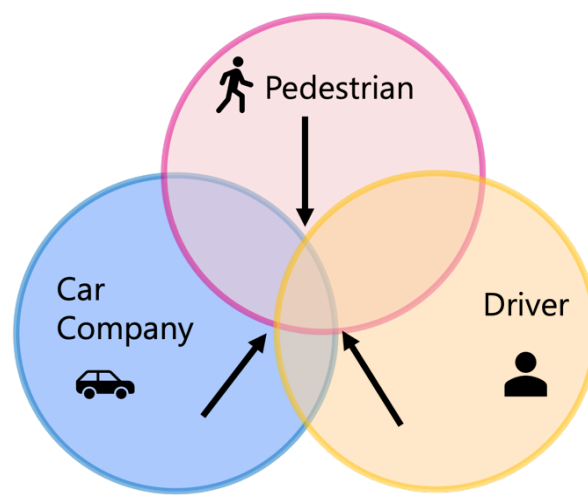

Figure 5. Different perspectives on the same incident.

b.  Many events viewed from one perspective for comprehensive understanding of a stakeholder.

Here is an example with two incidents addressed from the perspective of self-driving car companies:

**Incident 1:** A self-driving car suddenly started at an intersection while waiting for a red light. The car company claimed that the problem was later proven to be fabricated to avoid responsibility.

**Incident 2:** Google's self-driving car collided with a bus while driving. Google released a statement saying that the cause of the collision was that the algorithm programmed the bus to slow down and yield when there was room. The bus did not slow down, and Google said it would take partial responsibility.

Self-driving car companies as a representative stakeholder may have multiple operational targets such as technological innovation, benefiting society, and maximizing profits as shown in Figure 6. Their decision-making behavior, preferences and reactions may vary from one incident to another, depending on how, when and where the incident takes place.

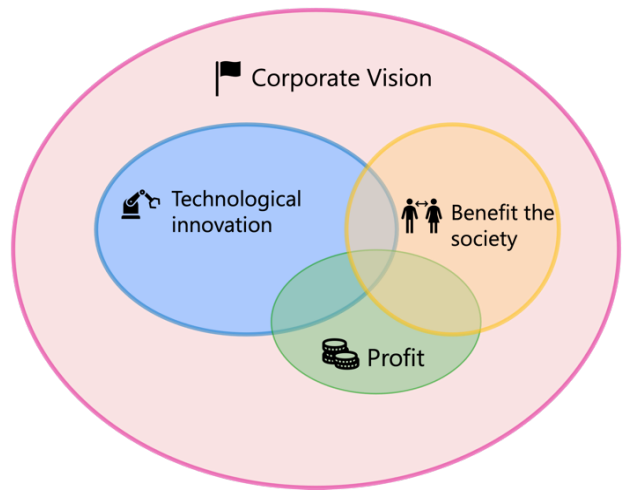

Figure 6. The perspective of a stakeholder with multiple operational targets.

c.  Many events viewed from many perspectives for holistic understanding and bias minimization.

Here is an example with two incidents addressed by different stakeholders:

**Incident 1:** An intoxicated, sleeping driver was driving a Tesla Model S with his hands off the wheel but still on autopilot. A highway patrol officer pulled him over and found the car in "assisted driving" mode. Stakeholders involved in this incident were police, driver, and lawyer.

**Incident 2:** Baidu, a giant Chinese technology company, introduced a self-driving car, the APOLLO Go, in Wuhan in July 2024. The self-driving car allowed passengers to realize a 24-hour on-demand and low-cost service. The incident aroused heated debate in social media. Stakeholders involved in this incident were passengers, car companies, and taxi drivers.

Both incidents are related to autonomous driving, but their key stakeholders are different with roles that are distinct and interactive as illustrated in Figure 7.

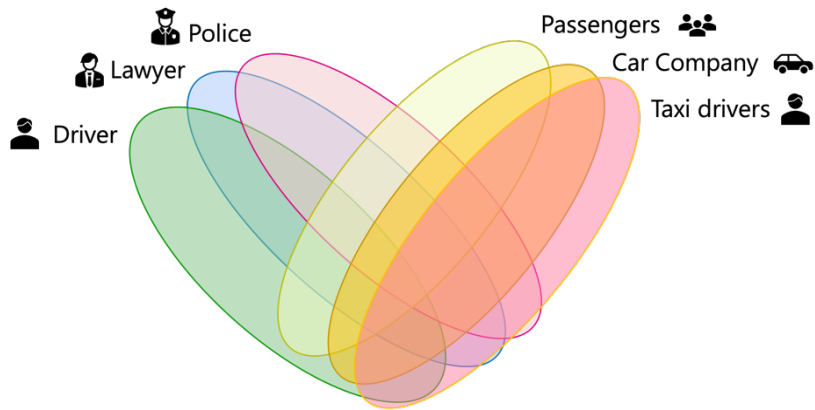

Figure 7. Multiple perspectives on Incident 1 (left side) and Incident 2 (right side).

Decomposing complex AI ethics events into representative scenarios lays the foundation for the knowledge driven design of data comics. Based on Venn diagrams demonstrated in Figure 6-8 that map out different stakeholders across various events, we can visualize key points of conflict and tension. This helps us extract the central narrative threads and transform raw news data into compelling comic stories, making ethical concepts more accessible, engaging, and interactive.

For scenario a), users exposed to engaging comics may become more willing to evaluate arguments representing three different perspectives and support finding common ground for reconciliation among the three parties. For scenario b), designing data comics to highlight the subtle changes in different contexts may help users to gain a comprehensive understanding of a shareholder. Finally, faced with multiple intricate relationships in scenario c), data comics are particularly useful in lowering the cognitive barrier for the average user and correcting cognitive biases for the knowledgeable user.

**5. FUTURE IMPLICATIONS OF THE ILA-DC FRAMEWORK**

Enhancing public understanding of AI ethics, from passive acceptance to active comprehension, is a critical prerequisite for ensuring that artificial intelligence technologies are deployed in society in a responsible and beneficial manner (Figure 8). As AI becomes an integral part of everyday life, the public is not only a user and subject of these technologies but also an indispensable stakeholder in the broader AI governance ecosystem (Machado et al. 2025). Therefore, understanding how AI systems operate, their associated ethical risks, and relevant governance mechanisms is essential for fostering public engagement, strengthening individual judgment, and enabling critical responses to AI system behavior.

The ILA-DC conceptual framework leverages visualization, narrative, and inclusivity to lower the knowledge barriers to understanding AI ethics, enabling individuals from diverse backgrounds to transition from confusion to reflection. This transformation fundamentally shifts the public's stance from passive acceptance of AI technologies toward a more agentic process of knowledge construction. Future research must address several core questions: How can visual storytelling be optimized to enhance ethical knowledge's comprehensibility and dissemination efficiency? Which cognitive pathways are most effective in helping the public internalize AI ethics as value judgments and behavioral choices? How can adaptive design support the development of culturally sensitive and contextually relevant AI ethics education mechanisms in diverse social contexts?

The ILA-DC conceptual framework provides a bridging mechanism between cognition and behavior. While understanding alone does not automatically lead to behavioral change, behavioral intentions are more likely to emerge when that understanding is grounded in concrete scenarios and embodied experience simulation (Kuhl and Beckmann 2012) (Figure 8). The presence of "ordinary person" characters in the comics – such as a user who inadvertently exposes his privacy or a student misjudged by an algorithm - serves as a proxy for the audience. These characters are relatable, emotionally engaging, and situated in specific contexts, enabling readers to analogize their behavior with the experiences of digital protagonists. Unlike approaches that merely present ethical issues, some data comics portray how the protagonist discovers, questions, and responds to unjust AI systems and seeks redress through appeals, public discourse, or collective action. These behavioral models are instructive, subtly encouraging readers to adopt similar strategies, such as adjusting privacy settings, critically selecting digital platforms, or participating in community advocacy. This mechanism strengthens the causal link between "my actions" and "societal outcomes," fostering a sense of ethical responsibility in digital behavior (Elliott and Spence 2017).

The ultimate societal impact of the ILA-DC framework goes beyond one-time information intake. While initial behavioral shifts may be triggered through contextual simulation and immediate feedback, significant educational influence lies in the stable internalization of these behavioral patterns as cognitive dispositions and value orientations, termed "learning habits" (Camic 1986; Gardner and Rebar 2019). This internalization is not an automatic process but requires the coordinated activation of multiple mechanisms. Within the ILA-DC framework, habit formation is conceptualized as a three-stage evolutionary system: structured input, reflective mechanism, and social embedding. At its core, this process transforms ethical behavior into cognitive automaticity and value adherence (Figure 8). Mere repetition is insufficient to create value adherence; instead, the emotional resonance, cognitive dissonance, and self-directed reconstruction in response to ethical dilemmas provide the deeper foundation for habit formation. By systematically building the data comics library of AI ethics incidents, the ILA-DC conceptual framework empowers readers not as passive recipients of ethical knowledge, but as moral agents who actively construct problems and reshape ethical judgments. Furthermore, the consolidation of any habit requires a social reference system (Pedwell 2017). The ILA-DC framework facilitates this by connecting AI ethics learning with broader societal discourse and community-based practices, fostering a co-learning social ecology. This aligns with inclusive learning analytics goals, which seek to cultivate a collective understanding of social issues. We aim to make moral judgment not merely an intentional act but a socially endorsed and habitual form of action, becoming an automatic choice aligned with personal identity and community belonging, ultimately contributing to a harmonious coexistence between humans and AI.

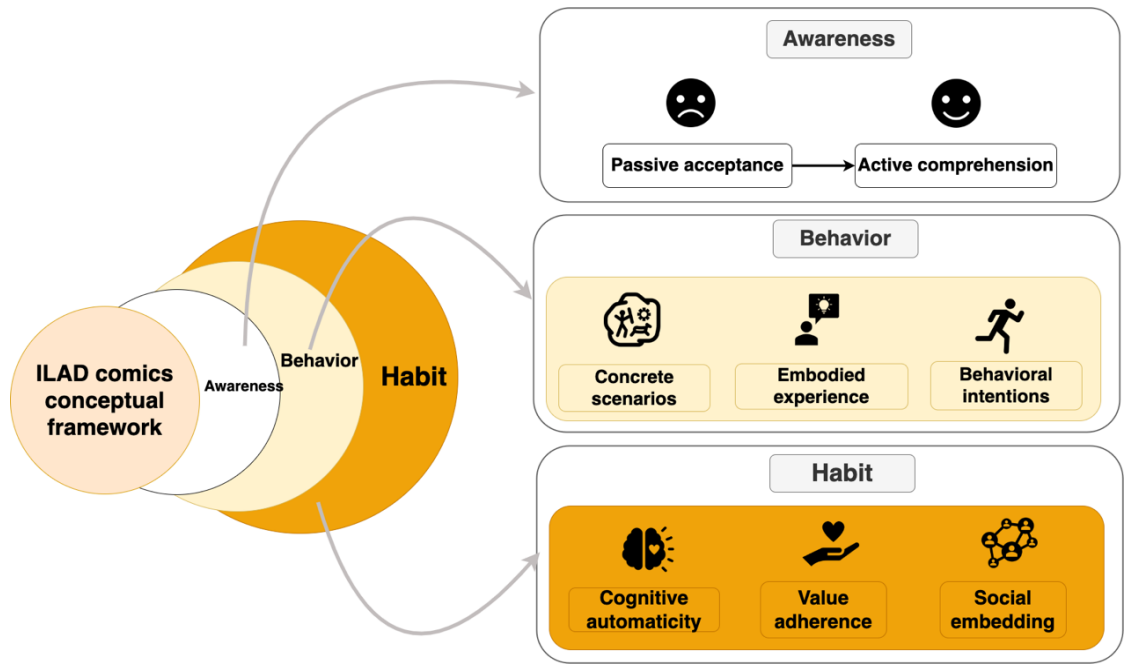

Figure 8. Future implications of the ILAD comics conceptual framework expectations

## 6 CONCLUSIONS

This paper presents a conceptual framework that explores how the inclusive learning analytics with embedded data comics can be used to handle complex AI ethical events. Instead of approaching the issue from traditional perspectives such as policymaking (Katsonis 2019) or public psychology (Rempel et al. 2018), we use real-world AI ethics incidents as the data source, decompose complex AI ethics events into representative scenarios for inclusive learning, which serves as the foundation for creating data comics. By embedding data comics in inclusive learning analytics, we seek to promote public understanding of AI ethics in an engaging and equitable way. In our follow-up work, we will empirically apply and evaluate the ILA-DC framework through workshops and user studies involving diverse publics. These studies will provide evidence on how inclusive learning analytics can foster ethical awareness and long-term learning behaviors in real-world contexts.

The development of AI technology is constantly advancing, and unknown risks are constantly emerging, indicating a growing need for long-term and lifelong learning. Our goal is not just to sensitize the public at a particular stage, but rather to cultivate public AI ethical literacy and encourage the formation of lasting, beneficial habits. This will help to achieve harmony between humans and AI.